\documentclass[conference]{IEEEtran}
\usepackage{pgfplots}
\usepackage{longtable}
\usepackage{lscape}
\usepackage{tabularx}
\usepackage{array}
\usepackage{enumitem,xcolor}
\usepackage{url}
\usepackage{cite}
\usepackage{amsmath,amssymb,amsfonts}
\usepackage{algorithmic}
\usepackage{graphicx}
\usepackage{textcomp}
\usepackage{xcolor}

\def\BibTeX{{\rm B\kern-.05em{\sc i\kern-.025em b}\kern-.08em
    T\kern-.1667em\lower.7ex\hbox{E}\kern-.125emX}}
    
\begin{document}
\title{A Comprehensive Survey of Aadhar \& Security Issues\
%{\footnotesize \textsuperscript{*}}
}
\author{\IEEEauthorblockN{Isha Pali, Lisa Krishania, Divya Chadha, Asmita Kandar}
\IEEEauthorblockA{\textit{Area of CSE} \\
\textit{NIIT University, Neemrana}\\ 
Rajasthan, India \\
(isha.pali, lisa.krishania, divya.chadha, asmita.kandar)@st.niituniversity.in} 
\and
\IEEEauthorblockN{Gaurav Varshney, Sneha Shukla}
\IEEEauthorblockA{\textit{Department of Computer Science Engineering} \\
\textit{Indian Institute of Technology, Jammu}\\
Jammu, INDIA \\
(gaurav.varshney, sshukla19.01.02)@gmail.com}

}

\maketitle

\begin{abstract}
The concept of Aadhaar came with the need for a unique identity for every individual.
%and the idea was first pitched in 2006 for BPL families. 
To implement this, the Indian government created the authority UIDAI to distribute and generate user identities for every individual based on their demographic and biometric data. After the implementation, came the security issues and challenges of Aadhaar and its authentication. So, our study focuses on the journey of Aadhaar from its history to the current condition. The paper also describes the authentication process, and the updates happened over time. We have also provided an analysis of the security attacks witnessed so far as well as the possible countermeasure and its classification. Our main aim is to cover all the security aspects related to Aadhaar to avoid possible security attacks. Also, we have included the current updates and news related to Aadhaar.

\end{abstract}


\section{Introduction}
\begin{figure}
\centering
\includegraphics[width=10cm,height=8cm]{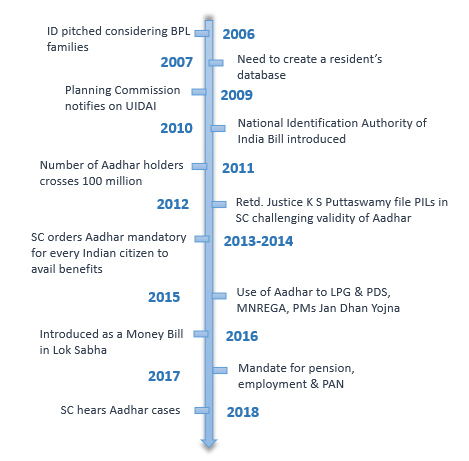}
\caption{Aadhaar Timeline}
\label{fig:timeline}
\end{figure}

As the name suggests, Aadhaar is the base of one’s identity in India. To provide a means of identifying an individual, the government of India has established UIDAI (Unique Identification Authority of  India)  in  2009  under the  Ministry of  Electronics and  Information  Technology,  following the provisions of Aadhaar Act 2016, to collect and manage the information of all the  Indian citizens.  Aadhaar consists of a  12-digit  unique number based on the user’s biometric and demographic data. The timeline of Aadhaar evolution is shown in Figure ~\ref{fig:timeline}.

Before the introduction of Aadhaar, the citizen identification in India was performed using passports, PAN cards, etc. but only a very few people have such cards and the more widely held identity cards such as ration card, NREGA cards, and voter ID cards had limited utility because of less portability and easy availability of fraudulent and spoofed cards, etc. This had created a lack of inclusion of such cards for identity verification in the country. After realizing the need for a unique trusted identity proof for every individual in the country, the government of India created Unique Identification Authority of India (UIDAI) to generate the Unique ID for everyone in the form of UID (Unique ID) linked with the iris scan and fingerprint as well as the demographic data of that individual. The card generated via UIDAI then got named as Aadhaar card. The word "Aadhaar" is a Hindi word that refers to a 'base' that verifies your identity.

\subsection{Aadhaar Features}\label{sec:subhead2}
There are various features of Aadhaar, which makes it a true and unique digital identity. Since the inclusion of Aadhaar has created a centralized system to manage the entire population, many services that require authentication are now possible with the use of Aadhaar card instead of using Proof of Address and Proof of Identity separately.
There are mainly five features of Aadhaar, which makes it more advantageous\cite{uidai}. 

\begin{enumerate}
    \item \textbf{Distinctive}: Biometric and demographic de-duplication provide the uniqueness of Aadhaar. Once a user has registered with his/her details with the biometric information, it will be stored in the UIDAI database. And if the resident enrolls again, he/she will be rejected as the data is already available and match exactly with the one present in the records. This makes Aadhaar a distinctive identity for every individual.  
    \item \textbf{Availability:} Since the Aadhaar authentication is available anywhere in India through online services. It gives a nationwide portability benefit. This is important as many people shifts from one place to another, which will not require the registration again.
    \item \textbf{Random Nature:} The Aadhaar numbers are generated randomly. A person who wishes to enroll must provide demographic data along with biometric information. It doesn’t require details about one’s caste, religion, income, health, etc.
    \item \textbf{Centrally managed Architecture:} The architecture of UID is scalable, which implies that the data is managed centrally and can be updated/created or authenticated from anywhere in the country. It is capable of handling 100 million authentications per day. \cite{uidai}
    \item \textbf{Technologies:} It uses open-source technologies, making it independent of specific computer hardware, storage, OS, database vendor, or any specific technology to scale. Since Aadhaar requires scalability the most, this feature will cater to it.
\end{enumerate}

\subsection{Aadhaar Benefits and Applications}\label{sec:subhead3}
There are several benefits of Aadhaar, which makes it more viable to use. Aadhaar has now been used in many services, which requires authentication.
The first benefit of Aadhaar is that it provides identities to all individuals irrespective of their caste, religion, income, or race. Since the registration requires the demographic data linked with the biometrics, there are fewer chances of fake identities and duplicates. Secondly, it enables financial inclusion and electronic transfer of benefits. Earlier, the people living in remote areas were neglected and were un-informed of the services available, or even if they knew, they were unable to use these services because of inadequate documents and details. With the presence of Aadhaar, they will also be identified and can avail services such as banking, loan, funds, sim-cards, etc. It also enhances the security of transactions by adding a level of security that requires authentication. This addition creates a governmental aspect between the transaction, which plays a safeguarding role in the eyes of individuals.  
All these benefits can be employed by making use of the applications, which require Aadhaar and its authentication services. Following are the top 10 Aadhaar applications \cite{raja}:
\begin{enumerate}
    \item Acquisition of Passport
    \item Opening Bank Accounts
    \item Digital Life Certificate
    \item Jan Dhan Yojana
    \item Disbursing Provident fund
    \item LPG subsidy
    \item Railway Reservation
    \item ATM security
    \item E-voting System
    \item Aadhaar e-KYC services
\end{enumerate}

This paper focuses on the authentication process and the security threats in Aadhaar. To cover that, the entire document is divided into seven sections. Section I describes the usage of Aadhaar, its features, benefits, and applications, followed by related work in Section II. Section III covers the Aadhaar authentication in detail. It talks about the Aadhaar model, the flow of Aadhaar, the analysis of UIDAI measures, the protection of user data, and possible measures against insider attacks. Section IV discusses the Aadhaar security and the potential security breaches and also explains the actual incidents that happened over the last two years. Section V presents the latest updates on Aadhaar taken from several accessible media sources. Section VI provides the conclusion of the entire study and the upcoming technologies such as Virtual IDs and Aadhaar coins. The last section covers the acknowledgment, followed by references.

\section{Related Work}\label{sec:relatedworks}
We did a literature survey to analyze the entire Aadhaar authentication process and its issues and challenges. 

In 2011, Vikas Sharma (ICDEOL) \cite{vikas} had presented the study on the benefits and challenges of Aadhaar’s invention. According to the author, Aadhaar has proven to be a potential benefit for underprivileged people who had to struggle to verify their identity at every level to get the benefits from government schemes and access the public distribution system.
%But along with the benefits, it faces a huge challenge to manage such an extensive database as mentioned by the author. 
The study showcases the 'Aadhaar-Unique Identification Model' along with the technological architecture of the unique identification system. 
%It is followed by the salient features and benefits of Aadhaar.
 The main challenge is the Implementation of the Aadhaar-UID System, which has been well pointed out by the author. The authors concluded that for the success of the Aadhaar project, it requires strong coordination between state governments and UIDAI at the top level and good communication between registrars and other intermediate agencies at the bottom level, for a good quality data enrolment and authentication. 

To further understand the need for Aadhaar, we referred to the report written by Chakrabarty et al. \cite{chakrabarty} in 2012. This document discusses how Aadhaar as a project was proposed in 2006 for “Unique ID for Below Poverty Line (BPL) families.” The prime aim of the unique ID was to provide the financially excluded ones with the services and the primary identity. 
%The Program Management Unit (PMU) was set up with a core team of experts who came up with KYR (Know Your Resident). 
This report talks about how UID verifies and authenticates Aadhaar in an online and cost-effective manner and how it can become an initial linkage for widespread financial inclusion in the country. The author then discussed the various advantages of UID in financial inclusions like Linking the UID number to a universal, accessible, and affordable micro-payment model that can transform the access the poor people have to bank services in the country, thus empowering every individual in India. 

Since our research area is Aadhaar Authentication as well as security, so we reviewed the papers written by other authors in this domain. For authentication, we referred to the Aadhaar Authentication API Specification released by UIDAI. The report \cite{api_1.5},\cite{api_1.6},\cite{api_2.0},\cite{api_2.5} describes the process of Aadhaar authentication wherein Aadhaar number, along with other attributes, are submitted to the Central Identities Data Repository (CIDR) for its verification. The Unique Identification Authority of India (UIDAI) provides an online service to support this process. Aadhaar authentication service only responds with a “yes/no,” and it does not return any personally identifiable information as a part of the response.
Further, it explains the multiple factors that are used for Aadhaar authentication, including biometric data, demographic data, OTP, PIN, possession of mobile, or combinations thereof and how the strength of authentication increases by adding multiple factors was discussed. The conclusion of the report states that the Aadhaar authentication provides a convenient or easy mechanism for all Aadhaar holders to establish their identity. It provides a platform for identity authentication and can be used to deliver services effectively to Aadhaar holders across the country. The report listed the changes in version \cite{api_2.5} from version \cite{api_1.5}. 

The second aspect covered in our study focuses on the security issues on the Aadhaar system. For that, we referred to several research papers to understand the privacy and other security issues of the Aadhaar system. In 2017 Singh et al.\cite{chal} have presented a review paper on Aadhaar card, its applications, and case studies related to data privacy and information loss. It further discusses the scope of Aadhaar and the benefits of linking Aadhaar card to various systems to avail of the services. To review these aspects of Aadhaar, the authors have presented their analysis based on several research articles, reports from leading newspapers, and Supreme Court verdicts. According to the findings in this paper, Aadhaar is one of the most noteworthy innovations towards digital transformation, which aids the citizens to use Aadhaar card as an identity in various areas such as railways, voting, PDS (public Distribution Systems), National Pension Scheme, etc. This paper highlights cases which require the initiative to solve security and privacy-related issues. The authors have also covered the Supreme Court observations on various Aadhaar issues, followed by a discourse on the ambiguities in the current system. 

Further, to understand the applications of Aadhaar in India, we reviewed the paper by Raja et al. \cite{raja} written in 2017. This paper presents a brief review of the Aadhaar card and discusses the scope and advantages of linking Aadhaar card to various systems (such as railways, health care, etc.). They also presented multiple cases in which Aadhaar card may pose security threats and harm the system.

The launch of the Aadhaar project was centered on the inter-operability of varied e-governance functionalities to confirm the optimum utilization of data, Communication, and Technology Infrastructure. %Towards this, the Government of India has now made Aadhaar card mandatory for many government applications and also has promoted Aadhaar enabled transactions.
 
In India, some of the Data Privacy and Protection laws are somewhat included under the sections of ‘IT Act (2008)’, and many are yet to be implemented and are under consideration. In the Aadhaar Act, some issues need to be understood. The paper also covers some issues with Aadhaar such as :
\begin{itemize}
\item Aadhaar is not a unique identity card; it’s simply a number. It does not contain any security features like PAN card, and Voter ID does.  
\item Aadhaar is used as proof of address. UIDAI does not even verify the address of the candidates. Still, Aadhaar has been accepted as a signal of address in the banking sector and telecommunication corporations.
\end{itemize}

In 2015, Agrawal et al. \cite{shweta}  looked into the possibilities of authentication using Aadhaar number, and it’s a database and proposed some ways to protect it. Being a single global identifier, Aadhaar is prone to privacy breaches and thus requires an alteration in design to make it safe. To know what is required for an individual application, a detailed analysis of the use case is needed. From a technical point of view, what they suggested: to get into several modern tools and techniques to make it up to date and provide a shield against run-time attacks. Moreover, they suggested an independent third party that can act as an auditor and keeper of the cryptographic keys used for encryption and decryption of data and transaction logs stored in the UIDAI systems. Insider attacks or leaks are the most significant threats. Under the administrative control of a third-party auditor, such malicious insider leaks can be prevented. Lastly, a robust policy framework to address the legal issues related to identification and authentication in a digital world was proposed. Security should be maintained on both the client-side and server-side to secure the database. The authors have suggested solutions to the existing attacks.

In 2017, A.K.R.S.Anusha and Dr. G. Rajkumar \cite{anusha} studied and verified the purpose, benefits, and the privacy and security issues faced during collection, storage, and transmission of Aadhaar details. The paper started with the description of Aadhaar's benefits and usage followed by the Privacy and Security Issues in the Aadhaar Life Cycle. This includes Vulnerabilities in Biometric Capture Devices, Private Players and Data Leakage, Cryptographic Algorithms, etc. The paper also discusses several active and passive attacks that may be targeted on the CIDR database. Further, it describes the laws and legislation related to these attacks. But the paper does not contain the possible solution to the problem. 

Another research work done by Arpana et al. \cite{aparna} in 2016 presented a study on Security Algorithms for Privacy Protection and Security in Aadhaar. The paper discusses various security measures that can be implemented to protect Aadhaar from insider attacks. In the Hadoop environment, the key requirements are authentication, authorization, data encryption, and security against various attacks. A new encryption technique to secure data in the HDFS environment is the combination of AES and Map Reduce.It performs encryption in parallel using AES-MR an Advanced Encryption standard based encryption using Map Reduce. To provide network security Quantum Cryptography and Biometric based Security solutions using BB84 protocol were suggested. The authors also explained the various proposed algorithms and security measure to implement on the Network level security Layer, Database level Security Layer and Application Level security Layer.

A summary of the literature has been drafted out in Table ~\ref{tab:related_works}. 

\begin{table*}[]
\caption{Related works}
\label{tab:related_works}
\begin{tabularx}{1\textwidth} { 
  | >{\raggedright\arraybackslash}X 
  | >{\raggedright\arraybackslash}X
  | >{\raggedright\arraybackslash}X
  | >{\raggedright\arraybackslash}X | }
 \hline
\textbf{References} & \textbf{Proposed} &\textbf{ Findings} & \textbf{Limitations} \\
\hline
Vikas Sharma(ICDEOL), 2011 \cite{vikas} & 
Study on the beneﬁts and challenges of Aadhaar’s invention. & Aadhaar, along with the beneﬁts, faces a huge challenge, such as the Implementation of Aadhaar-UID System. &  Implementation of the Aadhaar-UID System. \\

\hline
Singh et al. 2017\cite{chal} & Aadhaar card, its applications, and case studies related to data privacy and information loss. & Scope of Aadhaar and the beneﬁts of linking Aadhaar card to various systems to avail of the services.  &  Solutions to solve security and privacy-related issues. \\

\hline
Chakrabarty et al. 2012 \cite{chakrabarty} & How Aadhaar as a project was proposed in 2006 for “Unique ID for Below Poverty Line (BPL) families. & 
The various advantages of UID in ﬁnancial inclusions.
 &  Ways to authenticate yourself as an actual Aadhaar user.  \\

\hline
Raja et al. 2017\cite{raja} &  Aadhaar card, and discusses the scope and advantages of linking Aadhaar card to various systems. & The launch of this project was focused on the interoperability of various e-governance functionalities to ensure the optimal utilization of Information, Communication, and Technology Infrastructure. & Ways to remove loopholes in the existing system.  \\

\hline
Agrawal et al. 2015\cite{shweta} & Looked into the possibilities of authentication using Aadhaar number, and it's a database and suggested some ways to protect it.  & Insider attacks or leaks are the biggest threats. An independent third party that can act as an auditor to prevent malicious insider leaks & Under a third-party auditor, a solution to the minimization of attacks and leaks. \\
\hline
A.K.R.S.Anusha and Dr. G. Rajkumar 2017\cite{anusha} & Studied and veriﬁed the purpose, beneﬁts, and the privacy and security issues faced during collection, storage, and transmission of Aadhaar details. & Description of Aadhaar beneﬁts and usage followed by the Privacy and Security Issues in the Aadhaar Life Cycle. & Solutions about several active and passive attacks that may be targeted on the CIDR database.  \\
\hline
 \end{tabularx}
\end{table*}

\section{Aadhaar Authentication}\label{sec:auth}
Aadhaar authentication is the process wherein Aadhaar number with other attributes, which also include demographic data, biometrics, or a combination thereof, are submitted to the CIDR (Central Identities Data Repository ) for its verification. To understand the elements involved in Aadhaar Authentication in detail, refer to the official website of UIDAI \cite{uidai}.

\begin{figure*}
       \centering
        \includegraphics[width=17.cm, height=6cm]{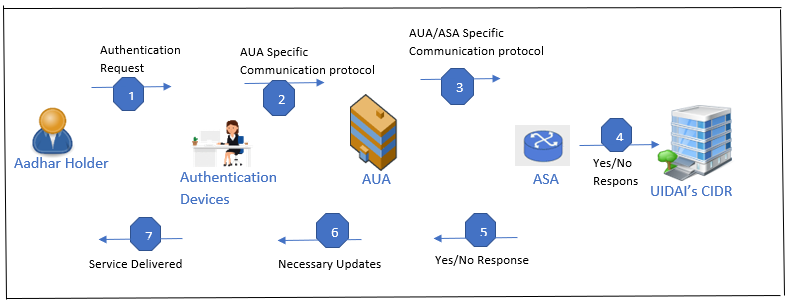}
    \caption{Aadhaar flow model}
    \label{fig:flow}
    \end{figure*}

\subsection{The Flow of Aadhaar Model:}\label{sec:subhead4}
Usage of open data format in XML and widely used stateless service such as HTTPS allows easy adoption and deployment of Aadhaar authentication.Following is the URL
format that was used for Aadhaar authentication service:
\begin{center}
    \textbf{\emph{https://auth.uidai.gov.in/2.5/public/uid[0]/uid[1]/asalk}} 
\end{center}
\begin{itemize}
    \item uid[0] and uid[1] are first 2 digits of Aadhaar Number when UID is used.
    \item asalk is a valid ASA license key which is received when device is registered successfully and it ensures that above mentioned URL is encoded to handle special characters.
\end{itemize}
The following are the key factors in Aadhaar authentication \cite{uidai}. The Figure~\ref{fig:flow}  explains the Aadhaar flow model. 

\begin{enumerate}
     \item Aadhaar number holder provides Aadhaar Number and a biometric detail which is unique to every person.
     The biometric detail was captured using a STQC certified and high-quality USB fingerprint scanner - MFS100.
    \begin{itemize}
        \item A data element "Bio" in PID XML contains an encrypted FMR(Fingerprint Image Record) in the case of registered devices.The FMR data is a fingerprint image packaged in ISO 19794-4 format, which could contain a lossy compressed image of type jpeg2000.
        \item Our application passes "UNKNOWN" in the posh attribute of the "Bio" data element in PID XML, as Aadhaar number holder can provide any random finger as their fingerprint.
    \end{itemize}
    \item Aadhaar authentication enabled application software installed on the device, calls the registered Device service to obtain PID, which is a base-64 encoded encrypted block.
    \item AUA application sends the encrypted PID block along with HMAC, Session key, Device attributes, etc. data to AUA server over mobile/broadband network using a standard and a secure protocol - HTTPS to support an end to end security and avoid request tampering and man-in-the-middle-attacks. This process includes the following steps:
     \begin{itemize}
        \item After forming PID XML, the SHA-256 hash of PID XML was computed and then it was encrypted with a dynamic session key using the AES-256 symmetric algorithm.
         \item Session key, in turn, was encrypted with 2048-bit UIDAI public key using an asymmetric algorithm(RSA). The session key was not stored anywhere except in memory and was used exactly once for every transaction.
        \item The encryption was done at the time of capture on the client and encoding(base-64) was done on the device which ensures no outsider to snoop and leak the data.
    \end{itemize}
    \item AUA server forms the final authentication XML input for API after validation, adds necessary headers(AUA specific wrapper XML with license key, digitally signs the XML for message integrity and non-repudiation purposes) and passes the request through ASA server to UIDAI CIDR.
    \begin{itemize}
        \item Digital certificate of class II, and the private key was used for digital signing.
        \item XML digital signature algorithm was used as recommended by W3C.
    \end{itemize}
    \item Aadhaar authentication server validates the input, decrypts data, validates decrypted data, does the matching, creates audit, and responds with a "yes/no" as part of the digitally signed response XML. If validation fails for some of the inputs in our application, then the following error was returned as a response:
    \begin{itemize}
        \item "300" : Biometric data did not match.
        \item "500" : Invalid encryption of session key.
        \item "502" : Invalid encryption of PID.
        \item "503" : Invalid encryption of Hmac.
        \item "540" : Invalid Auth XML version.
        \item "541" : Invalid PID XML version.
    \end{itemize}
    Other authentication errors, can be found at \cite{api_2.5}.
    \item Based on the response from the Aadhaar authentication server, AUA conducts the transaction.
\end{enumerate}
    Aadhaar system records all the authentication requests and their corresponding responses for audit purposes. UIDAI and AUA's digitally signs these authentication responses to validate the response integrity. In addition, few other attributes "ts", "info" within the API response are used to verify that the request was for legitimate Aadhaar user, if the request was having biometric factor when authentication was being done. Such self- verifiability of the response allows to trust third party applications. 

\subsection{Aadhaar's privacy concerns}\label{sec:subhead5}
The major concern about the Aadhaar scheme is the Aadhaar number. As it is a single unique identifier that must function across multiple domains, and hence it becomes not only electronically but publicly available but also in human-readable forms as well. This may lead to multiple breaches in privacy. The possible breach of privacy where an individual can be compromised in a setting such as in Aadhaar are as follows: 
\begin{enumerate}
    \item \textbf{Identification without consent using UID}: tracking of Aadhaar ID is done without the consent of respective identity, which may break authentication and confidentiality of the user.
    \item \textbf{Identification without consent using biometric data}: Aadhaar data can be illegally used to identify people with unauthorized use of biometrics such as fingerprint, iris scan, or facial photographs.
    \item \textbf{Illegal tracking of individuals}: records of an Aadhaar user with precise location, time, and context can be tracked down.
\end{enumerate}
The above attacks are much likely if the attacker can collude with the insider with access to various components of the Aadhaar System.

\subsection{Recommendation of UIDAI:}\label{sec:subhead7}
The UIDAI recommends that the AUAs should maintain the mapping between their domain-specific identifiers and the global Aadhaar number at their back-end. But UIDAI assumes that there cannot be any breach of privacy and thus does not maintain such mapping.
%Also, there are no safeguards or guidelines, either legal or technical, on how the Aadhaar number should be maintained and used securely, and how to prevent the Aadhaar number from being publicly available.

In the current scenario, the scheme requires the Aadhaar authentication in which the user needs to provide the Aadhaar number as a public identifier. Due to such weak provisions, identification without consent and correlation of identities remain as real possibilities. 
To solve the above-stated problem, \cite{shweta} suggested providing different local Aadhaar numbers for the different domains under one master identity embedded in a strong cryptographic method. This may prevent the connection among the domains if at all possible by allowing limited linkages across domain,uni-bidirectionally, or bidirectionally.

\begin{table*}[]
\caption{Comparison of Versions}
\label{tab:version_table}
\begin{tabularx}{1\textwidth} { 
  | >{\raggedright\arraybackslash}X 
  | >{\raggedright\arraybackslash}X
  | >{\raggedright\arraybackslash}X
  | >{\raggedright\arraybackslash}X | }
 \hline
\textbf{Version 1.5}\cite{api_1.5} & \textbf{Version 1.6}\cite{api_1.6} &\textbf{ Version 2.0}\cite{api_2.0} & V\textbf{version 2.5}\cite{api_2.5} \\
\hline
Meta element was inside PID block & Meta element outside PID block & New (2.0) Aadhaar holder consent for authentication & Incorporated BFD feature into core authentication\\
\hline
Meta element was optional & 
Meta element and attributes made mandatory.\newline
Structure changed to accommodate geo-location, public IP address.
 & Meta part has many additional attributes to support registered devices
 &  Usage of VID and UID token incorporated \\
\hline
Session is sent always. & Session key is synchronized with the allowance of advanced use case. & 
SSK scheme removed with introduction to registered devices.
 & 
 Session key must not be stored anywhere except in memory and should not be reused across transactions.  \\
\hline
“Txn” attribute , optional& “Txn” attribute made mandatory for better working of system & Info attribute of the response enhanced from version 2.0 to 3.0  &  Info attribute version changed to 4.0 from 3.0 \\
\hline
PID block is always XML. & 
Additional support for binary Protobuf format. & PID (Personal Identity Data) session key encryption changed to AES/GCM. & “Data” element contains “Pid” element which is a base-64 encoded encrypted block.  \\
\hline
 \end{tabularx}
\end{table*}

\subsection{Comparison of Versions}\label{sec:subhead10}
India’s Aadhaar project has increasing users and is getting more secured with every new version. So we have compared the four versions, mainly 1.5, 1.6, 2.0, and 2.5. Refer Table ~\ref{tab:version_table} for the comparison as mentioned above.

\section{Aadhaar Security}\label{sec:security}
The Aadhaar can be successful only if its implementation and use are secure at every step. For the safe use of this identity, UIDAI uses a well-protected authentication scheme. From the collection process to authentication, there exist many vulnerabilities that can be exploited by attackers.
To identify every individual uniquely, UIDAI has provided Biometric means to store the data corresponding to each user. The Biometric signature is used for an identity as well as the authentication process. But there are few requirements for biometric to work properly. It includes face, fingerprint, and Iris scan, which requires proper orientation and light for face detection, good quality of fingerprint. The use of contact lenses or spectacles can hinder the Iris authentication process. Also, there is a possibility of false positives in biometric authentication in a population of 1.3 billion, which requires a unique identity for everyone.
Biometric security is the main concern to make the Aadhaar system secure. As referred from \cite{dsci}, some security vulnerabilities in Aadhaar Biometric System includes:

\begin{enumerate}
    \item \textbf{Fake biometrics:} The attacker can create fake fingerprints, face, or IRIS images to steal the user's identity, and once they get the identity, the user cannot re-generate the genuine identity for themselves.
    \item \textbf{Backdoor Planting: }Instead of attacking every time, the attacker can plant a backdoor for next time. For e.g., circumventing the device by injecting a recorded image within the system input is much easier than attacking the device. 
   \item \textbf{Identity Stealing:} Once a user gets access to the storage, he can overwrite the legitimate user’s model/template with his/her own - in essence, stealing their identity.
    \item \textbf{Modification of Template:} To get a high verification score, no matter which image is conferred to the system, the feature sets, on verification, or in the templates can be modified.
    \item \textbf{Script to increase the verification score:}The attacker can put a Trojan horse program to some elements of the system, such as a matcher that always generates high verification scores. 
\end{enumerate}
User acceptance is another problem with Biometric Authentication since people don’t find sharing their personal information as a convenient mode for interacting with the government.%But people need to understand that e-Governance can be a possible way for each citizen of the country can get involved in certain government activities, to change existing chaotic and problematic government services.

\subsection{Security Challenges in a CIDR}\label{sec:subhead11}
The complete detail of the residents is stored in the UIDAI database - CIDR at a single location, which again can become a breeding ground for attacks. So protecting the central repository is a big challenge. The possible threats to the UID system are :
\begin{enumerate}
   \item\textbf{Unauthorized  writing: }If there aren’t any proper security measures then, the unauthorized user can write the files present in the database with malicious content. 
   \item \textbf{Data Sniffing: }This kind of threat occurs when someone intercepts the communication between CIDR and Registrar system and reads the data by sniffing data packets traveling in the channel.
   \item \textbf{Redirection to Fake server:}The normal user can be redirected to a fake server through phishing. This constitutes a privacy threat because the fake server can then access the user’s information.
   \item \textbf{Physical Breach:  }With the absence of physical security, the attacker can steal private keys to access the private data, which can be done by rewiring a circuit on the chip or using the probing pins to monitor data flows.
   \item \textbf{Cryptological attacks:} These attacks break the confidentiality of data transmitted as it directly targets the encryption algorithm. 
\end{enumerate}
Even though the CIDR system has been established by taking all the possible security measures, the intermediate agencies and registrars which capture and store the same data can become a weak link to be exploited.

\subsection{Security Challenges during Collection Phase}\label{sec:subhead12}
The data collection process involves the registrar who aggregates enrollment from intermediate agencies and sub-registrars and forwards it to CIDR. 
The security challenges here involves: 
\begin{enumerate}
\item \textbf{Compromization of account:} The admin can share its account to third parties, or it can be phished or attacked by viruses. 
\item \textbf{Insider Attack: }The enrolling registrars and agents can plant an insider attack for the sake of extra money or any kind of grudges with the government.  
\item \textbf{Insecure Communication Channel:} The details when entered through the application by sub-registrars, it is saved to the local database of and then get transferred to the registrar for submission to CIDR. If there is no secure channel present during this phase, the data can be altered or captured through the channel itself.  
\item \textbf{Outdated Software's:} The software used by the intermediate agencies who gather the details of users can be out of date, which poses a security vulnerability and can be exploited by malicious users.
\end{enumerate}
So, the collection process involves many security challenges, which can be avoided by secure communication between all the intermediate parties and the central repository CIDR.
\subsection{Security Challenges during transmission Phase
}\label{sec:subhead13}
After collecting information from the citizens, the next step is the validation of the details with the central database for de-duplication. This requires each intermediate agency to communicate with CIDR. Assuming that there will be a common application for every registrar office to communicate with each other, the transmission process attracts the maximum number of attackers. These include: 

\begin{enumerate} 
    \item \textbf{Man in the middle attack:} In this case an attacker gathers information and alter data and relay it.
    \item \textbf{Black hole attack:} A remote node acts as an intermediate node between two nodes, and packets are passed through this node. But this node makes all the packets disappear.
    \item \textbf{Bogus registration attack:} When someone registers himself with a bogus server and mask himself as genuine and gather all information.
    \item \textbf{Snooping:} This is similar to eavesdropping, but it only includes the usual observance of data flow.
    \item \textbf{Blackmail:}  It happens due to a lack of authenticity. It grants allowance to any node to corrupt data or information of other nodes.
\end{enumerate}
UIDAI must build multi-factor authentication to build a secure application and also need to take care of session timeouts. UIDAI must also have enough provisions to secure the endpoint of the transactional layer.

\subsection{Security Challenges during Storage Phase
}\label{sec:subhead14}
The Aadhaar data of 1.3 billion people, even if collected and transmitted by avoiding the possible threats, need to be stored securely, which requires a large size database. It is a huge challenge for the authorities to manage the security of such large sensitive information, which is segregated over various touchpoints across the country.
The possible threats to storage are:
\begin{enumerate}
    \item \textbf{Destructive Attacks:} There are many varieties of worms, viruses bots that can be used to attack the database to gather data, alter, delete data.
    \item \textbf{Incautious Behavior:}  The administrator can cause errors such as leaving the system unlocked, trip over wires, etc. which gives a loophole for an attacker to attack.
    \item \textbf{Infrastructural Error:} Server failure, systems with no physical security can also leave loopholes for attackers to get access to the database.
    \item \textbf{Original/Raw data:} The original data can easily be accessed as well as exploited, which can become a threat to the Aadhaar system. So data should always be kept in an encrypted or hashed form in the database.
     \item \textbf{Lack of proper access control mechanism:} Proper access rights are required to be given following the principle of least privilege and separation of duties. IT should implement regular control over the applications. And even when data is shared with other sources should be able to maintain confidentiality to avoid leakage.
\end{enumerate}
These were the possible security challenges in the Aadhaar system mentioned in several sources, such as the DSCI report. 

\begin{table*}[]
\caption{Security Breaches and classifications}
\label{tab:security}
\begin{tabularx}{1\textwidth} { 
   >{\hsize=.2\hsize\linewidth=\hsize}X 
   >{\hsize=.85\hsize\linewidth=\hsize}X
   >{\hsize=.85\hsize\linewidth=\hsize}X
   >{\hsize=.4\hsize\linewidth=\hsize}X}
 \hline
\textbf{Date} & \textbf{Breach} &\textbf{ Security Measures} & \textbf{Security Classification}  \\
\hline
18 Feb 2017 & 
The biometric information of some customers was used to activate additional SIM cards for other users by Reliance Jio employees in Madhya Pradesh.\cite{sec_breach} & Only trusted people should be kept for verification.\newline OTP based authentication must be enabled for issuance of SIM cards along with the biometric information of the user.
 & Espionage, Forgery, Deception, Malicious Third Party
\\
\hline
22 Feb 2017 & Third-party website leaked personal information which includes Aadhaar number, name, parents, gender, caste, signature and photo of 5-6 lakhs children.\cite{sec_breach} &
A proper access control mechanism should be maintained.Data should be stored in encrypted form.  Pseudonymization of data before storage.
 &Data Leakage.\\
\hline
24 Feb 2017 & Stored biometric information re-used for testing of authentication between January 11 to 17, 2017 by Axis Bank, Suvidhaa Infoserve, and eMadhura.\cite{sec_breach} & Strong multi-factor authentication must be implemented.
& Espionage, Misuse of Information, Malicious Third Party.
 \\
\hline
25 Feb 2017 & Aadhaar numbers and cards were being sold for less than the price of a cup of tea, with bulk purchases being even cheaper by data brokers.\cite{sec_breach} & 
Proper access control mechanism. Data at rest should be kept encrypted so that no one can read it even if the data gets leaked. 
& Data Leakage, Misuse of Aadhaar \\
\hline
30 Mar, 24 April 2017 & Aadhaar numbers were leaked by Swachh Bharat Mission, Ministry of Water and Sanitation (Central Govt.)\cite{sec_breach} &
Proper access control.Masking of data and only revealing the last 4-5 digits. & Third Party Leakage  \\
\hline
22 Apr, 23 April 2017
 & As a result of a programming error, Jharkhand Directorate of Social Security leaked the names, addresses, Aadhaar
the number and bank account details of beneficiaries of 1.6 million pensioners.\cite{sec_breach} & 
Version control should be done. Until the new version is made fully secured, the old version should be used.Once a new version securely starts performing, it should be brought to use. The new version should pass all security checks.
 & Insecure coding, Lack of proper testing\\
\hline
20 Apr 2017 & Without any form of authentication, Kerala SC/ST scholarship website leaked students’ information including their Aadhaar number, bank account number, and more.\cite{sec_breach} &
Proper access control mechanism.Pseudonymization of data. Implementation of Account lockout policy  & Data Leakage, Malicious Insider\\
\hline
25 Apr 2017 & Leakage of Aadhaar Number and a Photo
along with Name, Father’s Name, Gender, Age, Religion, Caste, City, and State by Pradhan Mantri Awas Yojna Site.\cite{sec_breach} & VAPT for all the third party websites using Aadhaar Data.
& Data Leakage, Malicious Insider.  \\
\hline
29 Apr 2017 & Leakage of various beneficiary details including name, ration card number, aadhaar card number, and more by Andhra Pradesh State Housing Corporation.\cite{sec_breach} & Pseudonymization of data.Proper Detect and Respond Mechanism in the websites & Data Leakage, Malicious Third Party, Malicious Insider.   \\
\hline
30 Apr 2017 & Research scholars data was leaked by Kerala university.\cite{sec_breach} & Pseudonymization of data.Implementation of SSL on the govt. websites using Aadhaar Data. (https websites)& Third Party Leakage
\\
\hline
1 May 2017 & 1,59,42,083 Aadhaar numbers were found leaked by National Social Pension Programmer.\cite{sec_breach} &  Pseudonymization of data. Support for remote administration and logging features. Creation of least-Privilege users and groups& Third Party Leakage
\\
\hline
25 Sept 2017 & 
Aadhaar database illegally accessed by an IIT graduate via an app Aadhaar eKYC by hacking into the server of an e-Hospital system. The eKYC routed all the requests through those servers.\cite{fp} & Third parties must have restricted access to data. Must enforce stronger API access control and Intrusion detection mechanisms. Regular Assessment of Security controls to prevent any attack on the e-hospital system. & Phishing
\\
\hline
4 Jan 2018 & The Tribune published a report - Rachna Khaira, a reporter of newspaper, “purchased”, for Rs500, a login ID and password that enabled her to access the database managed by UIDAI \cite{breach3} & Aadhaar data should be kept encrypted.Proper checking of logs so that no foreign access is there.proper intrusion detection systems in place to deter any anomalous behavior. Rate Limiting 
& Social Engineering 
 \\
\hline
5 Jan 2018 & Money was fraudulently withdrawn using customers' Aadhaar numbers from bank accounts reported by The Indian Express.\cite{breach2} & 
Implementation of Multi-factor Authentication. Training for the individuals in bank dealing at the user end.
 & Identity theft and Online Fraud
 \\
\hline
24 Mar 2018 & Data leak due to unsecured API endpoint informed by ZDNet \cite{unsecured_api} & 
 Better threat modeling before software and API releases.Rate Limiting on the API use. Implementation of a specific policy for securing Aadhaar Data.
& Insecure Coding.
 \\
 \hline
 31 Mar 2018 & Bank accounts opened using fake Aadhaar cards, for export-import business as reported by Hindustan times.\cite{breach7} & Proper access control mechanism. Data at rest should be kept encrypted so that no one can read it even if the data gets leaked. 
& Data Leakage, Misuse of Aadhaar \\
\hline
30 April 2018 & Andhra Pradesh government website leaked aadhaar data of 70 lakh children.\cite{breach8} & None & None \\
\hline 
1 February 2019 & Jharkhand Government online system leaked Aadhaar numbers along with the other personal details of thousands of workers.\cite{breach9} & None & None \\
\hline
19 February 2019 & Indane Gas website leaked millions of Aadhaar numbers, since part of website was indexed in Google, allowed anyone to bypass the login and gain access to the database.\cite{breach10} & None & None \\
\hline
15 April 2019 & Telugu Desam Party (TDP) hired IT company to develop Seva Mitra app, found to apparently store Aadhaar details of 7.82 crore users from Andhra Pradesh and Telangana.\cite{breach11} & None & None \\
\hline
20 September 2019 & The sensitive data like PAN card, Aadhaar and income tax details was leaked by Gujrat real estate authority website due to one of it's unprotected download URL.\cite{breach12} & None & None \\
\hline
\end{tabularx}
\end{table*}

\section{Security breaches and preventive measures}\label{sec:measures}
The Aadhaar ecosystem relies on three layers physical,data link and application layer. The user data used for Aadhaar enrollment is strongly encrypted with 2048 bit PKI which ensures secure transmission. However, infrastructure could be owned by third party agencies. Similarly, the application layer might be managed by non-UIDAI entities. While UIDAI requires for all contracting parties to put appropriate network security in place to ensure that their systems are protected from attack but, it is impossible to ensure systems-wide compliance. At the application layer, there is no nation-wide encryption policy to regulate data security.These applications doesn't ensure end-to-end encryption for user's personal information. It will become easier for an attacker to profile the user behaviour, if more number of such applications will link together Aadhaar numbers and user's personal information(in unencrypted form). This profile generation can be used against user to exploit them.
In our research, we have found out several security breaches that happened in the country which have been reported in various publications and reports. The growth of the country with the digital transformation brings the possibilities of cyber attacks as well. We have collected some security incidents that happened in  India (2017-18) and classified them in each cyber attack category, as mentioned above in Table ~\ref{tab:security}.

\subsection{Preventive measures for Security Attacks}\label{sec:preventive}
Aadhaar from being a random number which doesn't reveal anything about an individual data to other features like fingerprints, iris scan, address etc should be prevented from an intruder. To counter above mentioned strategic threats, we outline various preventive measures:
\begin{enumerate}
    \item \textbf{Use of secured communication channel}: VPN preferably SSL-VPN or the MPLS clouds can be deployed due to high sensitivity of the data during transmission phase.
    \item \textbf{Enabling HSTS (HTTP Strict Transport Security)}:The Man-in-middle agent makes user believe that an HTTPS connection is not available and that HTTP must be used. If user is not careful, then the connection with UIDAI servers will be in an un-encrypted form causing SSL stripping. If HSTS will be enabled before making connection request to CIDR/UIDAI servers then it will enforce HTTPS connection and will protect user against SSL stripping.
    \item \textbf{Secure testing of aadhaar enabled biometric applications}:The biometric template must not be stored , which allows user verification as it can lead to substitution attack where an attacker can overwrite legitimate user's template with it's own. Biometric Encryption can be implemented that securely binds a cryptographic key to a biometric, so  that neither the key nor the biometric can be retrieved from the stored template. The key is recreated only if the correct live biometric sample is presented on verification.
    \item \textbf{Isolating Black Hole Attacks}:Black-hole attacks exists on network layer where black-hole node aims to fool every node in the network that wants to communicate with another node by pretending that it always has the best path to the destination node. AODV integrated with a lightweight technique that uses timers and baiting can be used to isolate single and cooperative black hole attacks, preventing information to reach intruder.
    \item \textbf{Preventing insider attacks}: Within the UIDAI, the transaction logs are stored in encrypted form along with their decryption keys. The keys might be tampered and can be illegally accessed by recomputing HMAC without informing the database administrator. This is an insider attack which can be solved by an independent third party agency, being a keeper of cryptographic keys.
    \item \textbf{Hashing biometric data}: A sensitive information like biometric data of individuals can be protected by storing non-invertible hash of biometric data, so that information can't be spoofed. 
    \item \textbf{Deploying DNSSEC}: The CIDR is the centralized database where all the sensitive information of residents of the country is stored. Attackers can easily target UIDAI servers through DNS hijacking, attempting to trick them into revealing sensitive information, like fingerprint, face and voice templates.To prevent such scenarios DNSSEC (Domain Name System Security Extension) can be deployed at each of domain name infrastructures and on the DNS servers. In DNS servers the users' data are stored in unencrypted form. DNSSEC fixes this problem of unencrypted data by authenticating the origin  of the data. It help's DNS resolver to ensure that the received data is not tampered and is authentic. DNSSEC can be enabled using Cloudfare. 
    \item \textbf{Masking}: Memory cache at UIDAI servers contains all the frequently accessed information along with the key used for encrypting these sensitive data. A cache side-channel attack enables an attacker to recover the secret key depending on the accesses made by the user, thus deducing the encryption key used for securing the user's data. This method is invisible to user as it doesn't affect the ongoing cryptographic operations in aadhaar enrollment. Masking can be used as countermeasure to deal with all sort of side-channel attacks. It avoids manipulation of sensitive data (assume x here) directly by an attacker, rather manipulating a sharing of it. If \[ x_1, x_2, x_3, x_4,...... x_n \] are n number of shares a user data can have, then
    \newline
    \[ x= x_1 \oplus x_2 \oplus x_3 \oplus x_4 \oplus...... \oplus x_n \] In order to get the information, an attacker has to recover all the shares value to get the information about an aadhaar card holder.

\end{enumerate}
\section{Aadhaar-Enabled future applications}\label{sec:applications}
There are few major areas which need technology-fuelled structural reforms in our government, our business and our society at large.   
\begin{enumerate}
    \item \textbf{Election}: In order to eradicate voter fraud, a scheme to link voter ID with Aadhaar was launched in March 2015, since a person with registered biometric details can't vote again. Meshing Aadhaar with voter ID can speed up our sluggish voter enrollment process, only when the voting system becomes electronic. Once the Aadhaar linkage is in place, smartphone applications will allow every aspect of voting process- voter registration, address changes, polling booth information, even casting one's ballot- to be available to us on our smartphones.
    \newline
    This national ID program can be leveraged to allow citizens to vote using mobile phones along with a choice of polling stations. Instead of being restricted to vote at assigned poll booth, we must be able to cast our vote instantly at any polling station in any part of country using our voter ID linked with Aadhaar, obtained and verified through cloud. 
    \newline 
    It can serve as tagging device for new voters to be enrolled. The Aadhaar holders turning eighteen in a given year can be made eligible to vote, since the Aadhaar numbers are issued birth on wards, the system can automatically flag them.
     \item \textbf{Healthcare}: Telemedicine facilities across hospital chains has served as the first point of contact for patients, allowing doctors to directly interact with and diagnose patients sitting thousands of kilometers away. But, it is impossible to track any patient progression through the existing medical system as medical service providers have their on internal data storage systems, which operates independently of each other.
     \newline
     EMRS (Electronic medical record system) can be used as low cost and efficient patient management system, where multiple technology service providers will abide by set of interoperability guidelines, allowing users of platforms to pick vendors of their choice, as well making it easy to move from one service provider to another. The Aadhaar number will serve as "natural patient identifier" in such systems.  The Aadhaar backed EMRS will be a treasure trove of a 'big data' that can be mined using analytics to identify public health trends, collect statistical data, perform disease surveillance and detect epidemic outbreaks. 
     \newline
     But building an EMRS is cumbersome as it requires creation of new regulatory and legislation bodies, perhaps a new National Information Utility (NIU). Instead we can imagine a platform - National Health Information network (NHIN), that will act as central repository for all health data in the country. An asynchronous design which will allow enrollment of every single participant prior to the launch of NHIN. The government can use Aadhaar-linked NHIN to make health-related payments for individual in the similar way Aadhaar-linked bank accounts are used to outlay government payments. 
     \item \textbf{Education}: The RTE Act mandates that 25 percent of the total enrollment capacity should be set aside for economically deprived students, and free education should be given to such students up to elementary level. Some part of funding can be utilized to create "school vouchers" that can be accessible to deprived students to pay their education fees at the school of their on own choice. Aadhaar can be used in creation of a central registry and voucher-issuance platform for schools and students. Once student will register in system using Aadhaar, vouchers can be issued against the Aadhaar number of the students, and parent can use these vouchers to enroll their child at a school of their choice.
     \newline
     The Entrance examination fraud is another area of education sector where we need Aadhaar. For instance "engine-and-bogey" scam where the good student is asked to sit in the centre (engine) and other students will sit on either side (bogey) to copy in the exam. Such frauds can be eradicated using Aadhaar, if the students will be authenticated using Aadhaar before they enter the examination hall and seats will be allocated randomly once they enters the examination hall.
     \newline
     Another immediate application of Aadhaar is cracking fake resumes. The Digital locker initiative is encouraged by our government to de-materialize our educational degrees and skill certificates. Once combined with Aadhaar-based identification, person's education qualification is assured, thus "trust between employer and job seeker is established". The friction generated by the urge to authenticate transcript is eliminated, as no longer any individual have to get their transcripts attested and notarized. 
     \item \textbf{Electronic Toll}: Manual toll collection system augment traffic congestion, wastes fuel and doesn't serve the purpose for building a modern and high speed road network. In order to allow seamless flow of traffic during peak hours, electronic toll collection system could be implemented across the country. FASTag is a device that can be used to carry out cashless payments through a prepaid account linked to it at toll plaza - the equivalent of aadhaar for vehicles.
     \newline
     This RFID(Radio Frequency Identification) enable toll collection system is equivalent to telecom prepaid model. Tags distribution, the user registration, and the real-time transactions are three foundation of such toll systems. FASTag are low maintenance since vehicle owner needs the tags to be fixed to their windscreen only once, without worrying about the tag charging or battery changing overhead. The tags have installed antennas and circuit chip which is linked to their prepaid account, allowing transceivers installed at toll plaza to scan the QR code and the tag identification number, following which the funds will be debited from the account and the tag is topped up. Also, the SMS with the transaction details is sent to the tag owner. The transaction is also recorded in CES(Central Electronic Toll System) connecting all the participants including radio tag seller, user adding funds to prepaid tag and the toll manager.
     \newline
     Initially the toll collection using FASTags was around 20-22 percent. After mandating it went up to 60 percent. We are still far from our target of achieving 100 percent implementation. It could be realized with the help of proper analytics that can automate the collected data on real time basis, thus helping us to achieve the targeted financial returns, analyzing origin destination and assessing highway utilisation levels for different categories of vehicles.
     
\begin{table*}[]
\centering
\caption{Current Updates}
\label{tab:updates}
\begin{tabular}{|p{2.5cm}|p{14cm}|}
\hline 
\textbf{ Date} & \textbf{News} \\
\hline
Dec 18, 2018, & Aadhaar no longer mandatory for phones, banking as center proves changes to Law.\\
\hline
Dec 6, 2018, & All Aadhaar users may soon be able 
to opt-out, get their data deleted from UIDAI Servers.\\
\hline
Nov 4, 2018, & Aadhaar Authentication data cannot be stored for more than six months, and the court directed the government not to issue Aadhaar to illegal immigrants.  \\
\hline
Sept 28, 2018, & Supreme Court released a list of services where Aadhaar is not required.\\
\hline
Jan 11, 2019, & UIDAI introduces new two-layer security. UIDAI announced: \textbf{Virtual Authentication} through a 16-digit virtual identity number and limited the access available to a service provider. Also, UIDAI will provide "\textbf{Unique Tokens}" to ensure the uniqueness of beneficiaries.  \\
\hline
Feb 19, 2019, & Aadhaar leaks again: Indane Gas website, app leak data of 6.7 million subscribers    \\
\hline
March 1, 2019, & Cabinet approves Ordinance for voluntary use of Aadhaar for bank accounts, sim cards, etc.    \\
\hline
May 8, 2019, & DHFL Gets National Housing Bank’s Nod to Sell Stake in Aadhaar Housing Finance.    \\
\hline
May 24, 2019, & You can use VID in place of Aadhaar Card number for authentication and availing e-KYC services. VID is essential for services like checking Aadhaar Card biometrics details, downloading Masked Aadhaar Card, lock, and unlock Aadhaar Card.   \\
\hline
May 29, 2019, & Banks Can Use Aadhaar for KYC With Customer's Consent: RBI.    \\
\hline
June 12, 2019, & Cabinet Clears Bill Allowing Voluntary Use of Aadhaar as Proof for Opening Bank Accounts, Buying Phone Connections    \\
\hline 
September 19, 2019, & Aadhaar card holders doesn't need to submit any documents to update their details of email ID, mobile number, photograph, biometrics and gender on Aadhaar, only need to visit nearby Aadhaar Kendra. \\
\hline
October 28, 2019, & UIDAI launched Aadhaar chat bot to address the Aadhaar related grievances of users. \\
\hline
November 20, 2019, & AAP government plea High Court that Aadhaar should be optional, not mandatory for property registration and land mutation under the law. \\
\hline
December 3, 2019, & mAadhaar app new version launched allowing users to store soft copy of their Aadhaar in their smartphone.  \\
\hline
December 30, 2019, & The Ministry of Finance extended the PAN-Aadhaar linking to 31 March, 2020 for taxpayers. \\
\hline
January 1, 2020, & Around 28 Aadhaar Seva Kendras (ASK) was opened by UIDAI to support 114 standalone enrolments and update centers across the country.\\
\hline
January 24, 2020, & The Union Law Ministry approved Aadhaar card and Voter ID linking process. \\
\hline
February 3, 2020, & UIDAI permits Aadhaar card holders to update their demographic and biometric details free of cost. However, other service updates requires a  minimal charge of 50 rupees.\\
\hline
May 11, 2020, & Food Ministry extended the deadline of seeding Aadhaar with all the ration card/beneficiaries up to September 30, 2020. \\
\hline
\end{tabular}
\end{table*}

     \item \textbf{Energy and Power Consumption}: Our energy economy largely depends on traditional mode of power generation like coal, gas and to some extent nuclear power plants. Today's grid is one-way street that allows flow of electricity and information from power plant to consumer. They are incapable of handling supply complexities and inefficient in reversing the energy flow. In order to change the energy landscape of our country, we need two-way channels which allow integration of renewable source of energy into system and provide more information to consumers about their usage while monitoring their networks.
     \newline
     The Aadhaar linked power plant projects can be used to bring India's vast unmetered population on board as well help us to monitor the electricity load haul by each and every transformer, thus preventing the fraud. Smart grid can be one such application that uses an early warning system to reduce the down time by distributing the load evenly to the other transformers, when the load on a particular transformer is high. The identity verification could help the government to track the power generation and consumption pattern of power producers. Thus,smart grids can be used to meet the demand and supply of consumers with much greater accuracy in real time.
     \item \textbf{Judicial System}: Our India's judicial system lacks transparency, providing minimal information and less control over the progress of proceedings in the court. The routine tasks like documents generation, evidence collection and scheduling trials doesn't need a judge's intervention. Many technology-based judicial reforms have already been proposed to address these activities, but it didn't moved beyond planning stage. A new type of institution is needed that provide us with a centralized and technology-driven court operation platform.
     \newline
     National Judicial Network(NJN) can be one such institutional model to build and oversee such a platform, an entity delivering paperless justice. With Aadhaar and e-KYC we can digitize the documents and electronically tag the evidences to allow easy movement between courts. The documents and evidences can easily be uploaded in such a searchable system, allowing for better case tracking and performance management. The centralized platform will allow public to analyze some of the available data and stakeholders to enrol at their own pace, thus weeding out paper-based system and strengthening paperless judicial system.   
\end{enumerate}

\section{Latest Updates on Aadhaar}\label{sec:news}
The above Table~\ref{tab:updates} shows some latest news releases (2018-19) about Aadhaar gathered from the several sources \cite{news}.

\section{Conclusion}\label{sec:conclusion1}
Security and privacy are the main concern of Aadhaar as it is now used in many government applications and Aadhaar enabled transactions. We have gathered some breach incidences related to Aadhaar from 2011 to present from the online available public sources. We have classified each incident into a type of breach and also mentioned possible security measures for each. India’s Aadhaar project has more uses and is getting more secured with every new version. So, we have compared the four latest versions of Aadhaar and presented them in a tabular form \ref{tab:version_table}.

According to our findings, Third-party leakages and identity theft are the most common breaches that happened so far. The Aadhaar system itself is secure, but the third party websites of govt. Departments that are using the Aadhaar data or authentication service are not adequately secured. There is a strong need to have a proper security assessment of every govt — website before its launch. Implementation of SSL, Proper access control, Regular security assessment, Detect and response mechanism, etc. is a must before using any Aadhaar related data.
Security should be maintained on both the client-side and server-side to secure the database. The CIDR system has been established by taking all the possible security measures; the intermediate agencies and registrars which capture and store the same data can become a weak link to be exploited. The most important measure is to maintain a proper access control mechanism with the least privilege whenever some organization uses Aadhaar data.
UIDAI has introduced new two-layer security due to widespread concern over security breaches in Aadhaar. To eliminate data misuse, UIDAI has also introduced virtual IDs \cite{virtual_ids}. Instead of sharing the Aadhaar number, Virtual IDs can be used for verification with no fear of information leakage since a user can generate a VID and use it for a single purpose and can change it every time, making it impossible to trace one's identity or real Aadhaar information. Further to enhance the security of Aadhaar, the experts have come up with the idea of Aadhaar Coins- India’s own Cryptocurrency to make Aadhaar secure on which we may work in the future.

\section*{Acknowledgement}
The authors would like to thank the contributions of Mr. Anmol Singh, NIIT University for the work done in this paper.

\vspace{12pt}

\begin{thebibliography}{000}

\bibitem{vikas} Sharma, Vikas. "Aadhaar-a unique identification number: Opportunities and challenges ahead." Research Cell: An International Journal of Engineering Science 4.2 (2011): 169-176,April 2011.

\bibitem{chal}Raju, Raja Siddharth, Sukhdev Singh, and Kiran Khatter. "Aadhaar Card: Challenges and Impact on Digital Transformation." arXiv preprint arXiv:1708.05117, 2017.

\bibitem{chakrabarty} Chakrabarty, Nirmal Kumar. "UID (Aadhaar)—Its effect on financial inclusion." The Management Accountant 47.1 (2012): 35-37.


\bibitem{raja} Raju, Raja Siddharth, Sukhdev Singh, and Kiran Khatter. "Aadhaar Card: Challenges and Impact on Digital Transformation." arXiv preprint arXiv:1708.05117 (2017).


\bibitem{shweta} Sharma, Shweta Agrawal Subhashis Banerjee Subodh. "Privacy and security of Aadhaar: a computer science perspective." Economic and Political Weekly (2017).


\bibitem{anusha} A.K.R.S.Anusha, Dr.G.Rajkumar."International Journal for Research in Applied Science \& Engineering Technology (IJRASET)" ISSN: 2321-9653; IC Value: 45.98; SJ Impact Factor:6.887Volume 5 Issue VIII, August 2017.


\bibitem{aparna} Arpana Chaturvedi, Dr. Meenu Dave, Dr. Vinay Kumar,"Security Algorithms for Privacy Protection and Security in Aadhaar.",International Journal of Scientific Research in Computer Science, Engineering and Information Technology
© 2017 IJSRCSEIT | Volume 2 | Issue 6 | ISSN : 2456-3307
 2017.


\bibitem{api_1.5}UIDAI, Aadhaar Authentication API 1.5 Report, \url{https://www.scribd.com/document/72124822/Aadhaar-Authentication-API-1-5}


\bibitem{api_1.6} UIDAI,Aadhaar Authentication API 1.6 Report,
\url{https://authportal.uidai.gov.in/static/aadhaar_authentication_api_1_6.pdf}



\bibitem{api_2.0}UIDAI,Aadhaar Authentication API 2.0 Report \url{https://uidai.gov.in/images/FrontPageUpdates/aadhaar_authentication_api_2_0.pdf}


\bibitem{api_2.5}UIDAI,Aadhaar Authentication API 2.5 Report, 
\url{https://uidai.gov.in/images/resource/aadhaar_otp_request_api_2_5.pdf}


\bibitem{dsci}{Security and Privacy Challenges in the Unique Identification Number Project, Prepared by DSCI
With inputs from the Industry 21st January 2010,} \url{https://www.dsci.in/sites/default/files/documents/resource_centre/security%20and%20privacy%20challenges%20in%20the%20uidai%20project.pdf}


\bibitem{report} IANS, July 29, 2018, \url{https://www.thequint.com/hotwire-text/aadhaar-details-leaked-after-trai-chief-throws-breach-challenge}


\bibitem{breach2}THE QUINT,05 Jan 2018, \url{https://www.thequint.com/news/india/aadhaar-data-breach-glitches-data-security-compromised-earlier}


\bibitem{breach3}Anita Babu, 21 Jan 2018, https://www.theweek.in/theweek/specials/aadhaar-data-breach-proves-design-flaw.html


\bibitem{news} News18.com, December18 2018, \url{https://www.news18.com/newstopics/aadhaar.html}


\bibitem{uidai}UIDAI,Government of India,
\url{https://uidai.gov.in/ecosystem/authentication-ecosystem.html}


\bibitem{sec_breach} sflc/libraries,2017, 
url{https://sflc.in/uidai-aadhaar-breaches-and-leaks}


\bibitem{fp} Tech2 News Staff Sep 25,2018, \url{https://www.firstpost.com/tech/news-analysis/aadhaar-security-breaches-here-are-the-major-untoward-incidents-that-have-happened-with-aadhaar-and-what-was-actually-affected-4300349.html}


\bibitem{unsecured_api} The Wire Government,24 Mar2018,
\url{https://thewire.in/government/unsecure-utility-service-provider-is-leaking-aadhaar-details-says-report}


\bibitem{virtual_ids} Virtual Ids, 24 May 2019, \url{https://www.financialexpress.com/aadhaar-card/how-to-generate-virtual-id-for-aadhaar-card-check-process-and-benefits-watch-video/1587844/}


\bibitem{breach7} 
\url{https://www.hindustantimes.com/mumbai-news/40-bank-accounts-opened-using-forged-documents-in-mumbai-used-for-export-import-business/story-7JpH2wGBI4SwlIBnmhBdcL.html}


\bibitem{breach8}  Sukirti Dwivedi, April 30 2018,
\url{https://www.ndtv.com/india-news/despite-laws-no-action-against-government-agencies-displaying-aadhaar-data-1844747}


\bibitem{breach9} Gaurav Shukla, 1 Feb 2019,
\url{https://gadgets.ndtv.com/internet/news/aadhaar-leak-jharkhand-government-reportedly-exposed-details-of-thousands-of-workers-1986719}


\bibitem{breach10} Zack Whittaker, 19 Feb 2019,
\url{https://techcrunch.com/2019/02/18/aadhaar-indane-leak/}


\bibitem{breach11} Tech2 News Staff, 15 April 2019,
\url{https://www.firstpost.com/india/aadhaar-data-leak-details-of-7-82-cr-indians-from-ap-and-telangana-found-on-it-grids-database-6448961.html}


\bibitem{breach12} Yatti Soni, Inc42 Staff,20 Sep 2019,
\url{https://inc42.com/buzz/yet-another-data-leak-in-indian-government-database-exposes-multiple-citizen-ids/}


\end{thebibliography}
\end{document}